\catcode`\@11
\documentclass[11pt]{article}
\usepackage{dsfont}\usepackage{cite}
\usepackage{times} \usepackage{amssymb}
\usepackage{pifont}\usepackage{amsmath}\usepackage{amscd}
\usepackage{psfig}\usepackage{subfigure}
\usepackage[mathscr]{eucal}
\usepackage[all]{xy}
\xyoption{arc}
\xyoption{curve}
\numberwithin{equation}{section}
\newtheorem{exx}{Example}
\newenvironment{example}{\begin{exx}
\par\normalfont}{\null\hfill\end{exx}}

\newcommand\remove[1]{}
\newcommand\preprint[1]{\vspace{-1in}\vtop{\null\hfill
\parbox[t]{1.6in}{\small\sc #1\\\null}}
\vskip .5in\bigskip\normalfont}
\newdimen\arrayruleHwidth
\makeatletter
\def\Hline{\noalign{\ifnum0=`}\fi\hrule \@height 
\arrayruleHwidth \futurelet
\@tempa\@xhline}
\makeatother

\renewcommand\bar\overline
\newcommand\cf[4]{\bibitem{#1}{#2}.~{\it #3};~{#4}.}

\newcommand\ch[1]{\operatorname{ch}({#1})}

\def\cy{{Calabi-Yau}~}
\def\c{\mathcal{C}}
\def\C{{\mathds{C}}}
\def\cp#1{{\mathds{P}}^{#1}}

\newcommand\dercat[1]{\goth{D}^{\goth{b}}(#1)}

\newcommand\eg{{\slshape e.g.~}}
\def\eq#1{(\ref{#1})}

\newcommand\figref[1]{Figure~\ref{#1}}

\def\Q{{\mathds{Q}}}

\def\goth#1{{\mathfrak #1}}

\def\im{{\mathsf{Im}\thinspace}}
\def\ipull{\imath^{\star}}

\def\rt{\longrightarrow}

\def\td#1{\operatorname{Td}({#1})}

\def\vp{\varpi}

\def\sw{{Seiberg-Witten~}}
\newfont\sheafnt{rsfs10}
\def\sheaf#1{{\mbox{\sheafnt #1}}}

\newcommand\alg[1]{alg-geom/{#1}}

\newcommand\cqg[3]{Class. Quant. Grav.{\bf #1}~({#2})~{#3}}

\newcommand\hepth[1]{arXiv:hep-th/{#1}}

\newcommand\ijmpa[3]{Int. J. Mod. Phys {\bf A{#1}}~({#2})~{#3}}

\newcommand\jmp[3]{J. Math. Phys.{\bf {#1}}~({#2})~{#3}}

\newcommand\jhep[3]{JHEP {\bf {#1}}~({#2})~{#3}}

\newcommand\npb[3]{Nucl.~Phys. {\bf B{#1}}~({#2})~{#3}}

\newcommand\plb[3]{Phys. Lett. {\bf B{#1}}~({#2})~{#3}}
\newcommand\prd[3]{Phys. Rev. {\bf D{#1}}~({#2})~{#3}}
\newcommand\prl[3]{Phys. Rev. Lett. {#1}~({#2})~{#3}}

\begin{document}
\title{\preprint{hep-th/0409090}  Strings, Junctions and Stability }
\author{ 
Avijit Mukherjee
\thanks{avijit@maths.adelaide.edu.au} \\ 
\small Department of Pure Mathematics \\
\small \& Department of Physics\\
\small  University of Adelaide \\
\small  Adelaide, SA 5005. Australia\\  
\and 
Subir Mukhopadhyay
\thanks{subir@physics.umass.edu} \\ 
\small Department of Physics \\
\small University of Massachusetts \\
\small  Amherst, MA 01003-4525\\  
\small  USA.
\and 
Koushik Ray \thanks{koushik@iacs.res.in} \\
\small Department of Theoretical Physics\\
\small  Indian Association for
 the Cultivation of Science\\
\small  2A \& B Raja S~C~Mullick Road,\\ 
\small  Calcutta 700~032. India.
}
\date{}
\maketitle
\vfil
\begin{abstract}
\noindent 
Identification of string junction states of pure $SU(2)$
Seiberg-Witten theory as B-branes wrapped
on a \cy manifold in the geometric engineering limit is discussed. 
The wrapped branes are known to
correspond to objects in the bounded derived category of coherent sheaves on the
projective line $\cp{1}$
in this limit. We identify the pronged strings with triangles in
the underlying triangulated category using $\Pi$-stability. The spiral strings in the
weak coupling region are interpreted as certain projective resolutions of the
invertible sheaves. We discuss transitions between the spiral strings and
junctions using the grade introduced for $\Pi$-stability through the central charges
of the corresponding objects. 
\end{abstract}
\thispagestyle{empty}
\section{Introduction}
\label{sec:intro}
String junctions are states in the type-IIB string theory 
formed by more than two strings
joining at a vertex conserving certain charges. Their 
existence is crucial for the concinnity 
of duality symmetries of string theories \cite{jhs}. 
BPS string junctions preserve a quarter
of the original supersymmetry (SUSY). String junction 
solutions have been obtained and used
in a variety of situations in F-theory 
\cite{asen1,mns}, M-theory
\cite{jhs,witten1} and string theories in ten
and lower dimensions 
\cite{alok1,subir1,alok2,alok3,alok4,alok5,shib}. 
In this article we concern ourselves with such states  
and phenomena related to them in
supersymmetric field theories, realised by D-branes in 
string theories \cite{asen3,sunil,bf1,bf2}. 

The objects of interest in this article are the 
half-BPS dyonic states in 
a four dimensional $\mathsf{N}=2$ super-Yang-Mills 
theory (SYM) with gauge group $SU(2)$, known as 
the \sw theory \cite{sw}.  
It can be viewed as the
gauge theory on the world-volume of a D3-brane in the 
type-IIB string theory in the
presence of background 7-branes.
The spectrum of this theory contains dyonic states, 
carrying electric as well as magnetic
charges under the gauge group, identified with appropriate 
RR-charges of the D3- and the
7-branes in the type-IIB theory, in which the 
\sw gauge theory is embedded.
Certain configurations of these dyons are interpreted 
as junctions in which at least
three strings join at a vertex, while the other ends of the strings 
end of the D3-brane or the 7-branes. 
At the vertex the electric and the magnetic charges of 
the strings are balanced, as in 
electrical circuits, rendering the configuration neutral. 

In a T-dual description as states in the type-IIA theory compactified on a
\cy threefold, ${\mathcal{M}}$, these states
arise as even-dimensional branes wrapping holomorphic cycles in ${\mathcal{M}}$.
The \sw theory, thus, has another description in terms of wrapped even-dimensional
branes in the type-IIA theory. Within the scope of topological theories, these are
given by branes in the topological B-model, alias B-branes, wrapping holomorphic
cycles in ${\mathcal{M}}$. A way of obtaining the \sw theory, known as
\emph{geometric engineering} \cite{kkv},
produces gauge groups by compactifying the type-IIA
theory on a suitable \cy manifold, then zooming into regions in its moduli
space where the manifold degenerates and gravity decouples.  
The resulting gauge group  depends on 
the specific manner of degeneration. Thus, the string junctions of the \sw theory 
are interpreted as wrapped B-branes on some suitable degenerate \cy manifold. 

Now, B-branes wrapping holomorphic cycles in ${\mathcal{M}}$ 
are described as objects in $\dercat{\mathcal{M}}$,  the
bounded derived category of coherent sheaves on ${\mathcal{M}}$ \cite{doug1}.  
This description is valid when ${\mathcal{M}}$ is sufficiently large, 
so much so that geometric notions
make sense. The branes that are stable are described in 
$\dercat{\mathcal{M}}$ by introducing a grade
guided by the preservation of SUSY. 
The corresponding criterion for stability, that is, between two configurations of
branes the one with a lower grade is more stable, is called $\Pi$-stability.
Altering the coupling in the \sw theory corresponds to perambulating the
K\"ahler-moduli space of ${\mathcal{M}}$, whereby the set of $\Pi$-stable objects in 
$\dercat{\mathcal{M}}$ 
mutate giving way to different sets of stable objects in different regions.
In the limit appropriate for deploying
geometric engineering techniques, the spectrum of dyonic states in the \sw theory
with $N=2$ SUSY arise from
the $\Pi$-stable B-branes \cite{ak} in the weak as well as in the strong coupling
regimes. 

Dyons in the \sw
gauge theory on the world-volume of a D3-brane probe 
are interpreted as string junctions in the  type-IIB theory 
in which the gauge theory is embedded
\cite{asen2,bf1}.
Moreover, as one perambulates the type-IIB moduli space,
the junctions go through transitions producing 
other junctions or strings. The latter, before ending on a 7-brane,
spiral around the marginal stability line, the line of demarkation
between the weak and strong coupling regions of the gauge theory 
\cite{bf2}. Upon interpreting the dyons as the $\Pi$-stable objects 
in $\dercat{\mathcal{M}}$, therefore,
it is expected that such phenomena can be interpreted 
using $\Pi$-stability. We show that this is indeed the case. 

In the abelian category underlying the derived category 
$\dercat{\cal M}$, a dyon may be viewed as a bound state of others, 
at  certain points in the moduli space, 
according to certain projective resolutions provided by short exact sequences of
sheaves on $\cal M$, pulled back from the projective line $\cp{1}$. 
This corresponds to a junction with three prongs, one lying in the weak coupling
region, and two in the other, with the vertex lying on the
marginal stability line, alluded to above. This corresponds, then, to replacing a
sheaf by a projective resolution, in which the original sheaf is stable in the weak
coupling region, while the two sheaves constituting the resolution are stable in the
strong coupling region enclosed by the MS-line.
In the corresponding triangulated category the short exact sequences correspond to
distinguished triangles. Thus, a three-pronged junction, possibly with multiple
strings per prong, corresponds to a distinguished triangle in the triangulated
category or a short exact sequence in the abelian category. The same dyon may, at
some other point in the moduli space, be viewed as a string that starts on the
D3-brane and ends on either of the 7-branes, staying always outside the line of
marginal stability and possibly going around it several times \cite{bf1}. We refer to
these as \emph{spiral strings}. We construct the projective 
resolutions of the sheaves in the
abelian category that correspond to these spiral strings in an iterative fashion,
corresponding to the invertible sheaves of rank $n>1$ on $\cp{1}$. In the derived category,
then, one can use either of the resolutions to limn the dyons corresponding to a
particular sheaf. We show, however, that there are preferred regions in the moduli
space corresponding to each of the resolutions, determined by the grade of
$\Pi$-stability. Passing from one such region to
another correspond to transitions between strings and junctions \cite{bf2}.

The organisation of the article is as follows. In the next section we recall a few
notions related to $\Pi$-stability and fix notations. In \S\ref{sec:shvdy} we briefly
review the identification of stable sheaves on $\cp{1}$ as \sw dyons discovered
earlier \cite{ak}. In \S\ref{sec:spiral} we obtain the projective resolutions that
correspond to the spiral strings illustrating them with a few examples. We then
proceed to discuss the transitions between the spiral strings and the junctions
before concluding.
\section{$\Pi$-stability \& dyons}
\label{sec:review}
Let us start by briefly recalling the notion of $\Pi$-stability. An object of the derived
category $\dercat{\mathcal{M}}$ is a triangle from the underlying triangulated category, 
\begin{equation}\label{abca}
\xymatrix{
A\ar[r] & B\ar[r] & C \ar[r] & A[1]
}, 
\end{equation} 
where $A$, $B$ and $C$ are complexes of coherent sheaves on ${\mathcal{M}}$,  
and $[n]$ denotes the
shift functor shifting a complex toward the left by $n$ degrees; so $[1]$ shifts a
complex by unit degree.
This in turn corresponds to short exact sequences of complexes in the underlying
Abelian category,
\begin{equation} \label{abc}
\xymatrix{
0\ar[r] & A \ar[r] & B\ar[r]&C\ar[r]& 0},
\end{equation} 
which can also be thought of as a projective resolution of the object $C$.
In order to define a grade on the derived category
one first
defines a map from the Grothendieck group of $\cal M$ to the complex plane
$Z: K(\cal M)\rt\C$, called the \emph{central charge}
\cite{bridgeland,doug1,doug2,doug3}. The central
charge  of an object $E$  of $\dercat{M}$  is given by 
\begin{equation}
\label{ze:ch}
Z({E}) = \int_{\mathcal{M}} e^{-(B+iJ)} \ch{{E}}\sqrt{\td{{\mathcal{M}}}},
\end{equation} 
at the large volume asymptote, with 
$B+iJ$ denoting the complexified
K\"ahler class of $\mathcal{M}$. 
The phase of the central charge defines a \emph{grade}, $\varphi (E)$, 
for every object $E$ as
\begin{equation} \label{faz}
\varphi ({E}) = -\frac{1}{\pi}\,\arg Z({E}),\quad -2<\varphi (E) \leq 0.
\end{equation} 
The magnitude $|Z({E})|$ furnishes the mass of the state. 
The criterion of $\Pi$-stability is formulated in terms of the grade $\varphi$.
A physically meaningful description of branes corresponding to \emph{all} the objects
in the triangle \eq{abca} or the exact sequence \eq{abc} exists if the grades satisfy 
\begin{equation}
\label{piphi}
\varphi (B) - \varphi (A) = 1,\quad\text{with}\quad \varphi (C) = \varphi(B). 
\end{equation} 
This embodies the condition of alignment of  charges of the branes and guarantees
that the corresponding solution is a BPS state in the theory preserving
$\mathsf{N}=1$ SUSY. 
A brane corresponding to the object $C$ in \eq{abca} is said to be \emph{stable}, 
\emph{marginally stable} or \emph{unstable} with respect to decaying into $A$ and
$B$ if $\varphi (B) - \varphi(A) \lessgtr 1$, respectively. The region of the moduli
space where it is marginally stable is called
a \emph{marginal stability locus}. 
Thus, to start with, 
we have a good description of topological branes on the marginal stability locus. Once the
marginal stability locus is ascertained, the brane corresponding to the object $C$ is
stable or unstable with respect to decaying into $A$ and $B$, as one
moves off this locus in the K\"ahler moduli space,  if 
$\varphi (B) - \varphi(A) - 1$ is negative or positive, respectively. 
This criterion is called $\Pi$-stability. In the sequel we often refer to $A$ and $B$
as \emph{the components} of $C$, in the context of a specific triangle. 

Let us now briefly summarise  the interpretation of  \sw dyons as $\Pi$-stable 
objects in $\dercat{\mathcal{M}}$ in
the geometric engineering limit \cite{ak}
thereby identifying the string junctions as objects
in $\dercat{\mathcal{M}}$. We then check how the analysis of
$\Pi$-stability reproduces the transitions between strings and junctions. 

The pure $SU(2)$ \sw theory is geometrically engineered by zooming in around the
principal component of the discriminant locus of the mirror of a \cy manifold ${\mathcal{M}}$,
which is a degree eight hypersurface in the resolution of $\cp{4}(2,2,2,1,1)$. The
sole compact part left of ${\mathcal{M}}$ in this limit is a curve $\c$ isomorphic to
the complex projective line $\cp{1}$, 
embedded in ${\mathcal{M}}$; $\imath : \c\hookrightarrow {\mathcal{M}}$, $\c\cong\cp{1}$.
The states which become massless on the discriminant locus
correspond to objects in $\dercat{\mathcal{M}}$ that are pull-back of 
objects of $\dercat{\c}$, indicated by
$\imath^{\star}$. Thus, for the purpose of studying the gauge theory, it suffices to
consider $\dercat{\c}$, whose objects are constructed out of the structure sheaf
$\sheaf{O}_{\c}$ and the torsion sheaf $\sheaf{O}_p$, concentrated at a point $p$
of the projective line, $\c$. The spectrum of the gauge theory is envisaged as the
category of stable objects in $\dercat{\cal M}$. 

In order to study $\Pi$-stability, one first identifies the pull-back of 
$\sheaf{O}_{\c}$ and
$\sheaf{O}_p$ as a D6-brane and a D4-brane, respectively, and associates the
corresponding central charges to the corresponding 
complexes of coherent sheaves on ${\mathcal{M}}$. Thus in the large volume asymptotic
\begin{equation}\label{period}
\vp = Z(\imath^{\star}\sheaf{O}_p), \qquad \vp_D = Z(\imath^{\star}\sheaf{O}_{\c}),
\end{equation} 
where the periods $\vp$ and $\vp_D$ are the two solutions of the Picard-Fuchs equation 
of ${\mathcal{M}}$ in the
limit under consideration. This entails the identification 
$\sheaf{O}_{\mathcal{M}} = \ipull\sheaf{O}_{\c}$. 
This provides the central charge map from $\dercat{\mathcal{M}}$ 
to the Grothendieck group of ${\mathcal{M}}$, alluded to earlier, 
the latter being generated in the present case by the zeroth and the second cohomology
groups of $\c$, $H^0(\c,\Q)$ and $H^2(\c,\Q)$, respectively. 
The asymptotic expression of the central charge is used to fix the periods $\vp$ and
$\vp_D$ in \eq{period} up to a constant. 
Furthermore, the monodromies around the points $z=\pm 1$ and $z=\infty$ fix 
the periods $\vp$ and $\vp_D$ to the \sw periods \cite{sw,bf2,bilal}, 
up to an overall constant \cite{ak}. Generally, 
the central charge of an object ${E}$ of $\goth{C}({\mathcal{M}})$, the category of
complexes underlying the derived category $\dercat{\cal M}$, can be written as 
\begin{equation} \label{zedE}
Z({E}) = m \vp_D + n \vp, 
\end{equation} 
where the integers 
$m$ and $n$ are, respectively, the rank $\big(\dim H^0(E)\big)$ and the degree $\big(\dim
H^2(E)\big)$ 
of $E$ defined through
the Chern character map. We can therefore denote the state corresponding to
${E}$ as $(m,n)$. The sheaves $\ipull\sheaf{O}_{\c}$ and $\ipull\sheaf{O}_p$ can
then be identified as the states $(1,0)$ and $(0,1)$ respectively.
The grade of $E$ is defined by \eq{faz} as the phase of the central
charge \eq{zedE}. These states are identified as the dyons in the \sw theory, with
electric and magnetic charges $n$ and $m$, respectively. 
\section{Stable sheaves and dyons}
\label{sec:shvdy}
The spectrum of dyons in the \sw theory consists in the $\Pi$-stable objects of
$\dercat{\cal M}$, obtained from $\dercat{\cal C}$ under the pull-back $\ipull$.
Thus, in order to find the dyons we need to analyse $\Pi$-stability of the 
sheaves on $\cal C$ and their complexes thereof. The dyons are identified as
$(m,n)$-strings or string junctions in the type-IIB theory \cite{bf1}. 
As mentioned above, an analysis of $\Pi$-stability involves two steps. First, the
marginal stability locus is to be obtained. Then stability as one goes off this locus is
to be checked. 

Thus, in order to study $\Pi$-stability of the dyons in the 
\sw theory we first need
to ascertain the marginal stability locus that, in the 
case at hand, is but a line,
which we refer to as the \emph{MS-line} in the sequel. 
Let us consider the objects in \eq{abc}. Let the central charges
of the branes involved be $Z(A) = m_A \vp_D +
n_A\vp$, $Z(B) = m_B\vp_D + n_B\vp$ and $Z(C)=m_{C}\vp_D + n_{C}\vp$, 
respectively, with
$m_{C} = m_B - m_A$ and $n_{C} = n_B - n_A$. 
Then, the MS-line is given by 
\begin{equation}
\begin{split}
1 &= \varphi(B) - \varphi (A) \\
&= -\frac{1}{\pi} \arg\left(\frac{m_B \vp_D/\vp +n_B}{m_A 
\vp_D/\vp+n_A}\right),
\end{split} 
\end{equation} 
which is satisfied only if $\vp_D/\vp$ is a real number. 
Thus, the MS-line is given by 
\begin{equation}
\arg (\vp_D/\vp)  = 0.
\end{equation} 
The curve in the $z$-plane is homotopic to a circle 
(see \eg \cite{matone} and references therein 
for works related to MS-line) separating the strong and weak
coupling regions of the \sw theory, indicated by 
$\goth{S}$ and $\goth{W}$, respectively, in
\figref{msl}. The MS-line goes through the points $z=\pm 1$.
The two 7-branes and the D3-brane of the type-IIB probe picture are all transverse
to the $z$-plane. They are thus represented by points in this plane. The 7-branes are
situated at $z=\pm 1$, while the D3-brane moves around in the plane. 
\begin{figure}[h]
$$\xy
0,/r3pc/:
*=+\dir{*}*+!UR{\scriptstyle z=-1}*+!DR{\ipull\sheaf{I}_p};
,p+(.2,1.3)*+{z}*\frm<2pt>{,},
,p+(2,0)*\dir{*}="o",*+!UL{\scriptstyle z=1}*+!DL{\sheaf{O}_{\mathcal{M}}}
,{\ellipse_{--}}
,{\ellipse^{--}}
,p+(1.5,.6)*{ \goth{S}}; +(1.7,1)*+{\goth{W}};
\endxy
$$\caption{MS-line separating $\goth{S}$ and $\goth{W}$}
\label{msl}
\end{figure}
In order to test the $\Pi$-stability of the different objects of 
$\dercat{\cal M}$ off
the MS-line, $\arg (\vp_D/\vp)$ is defined as a continuous real-valued 
function of $z$ \cite{ak}. Using \eq{ze:ch}, $\arg (\vp_D/\vp)$
is fixed to be $\pi/2$ at the large volume asymptote. 
It increases through $\pi$ on the
MS-line. From the expression of the central charge 
in terms of the periods, \eq{zedE}, then, we
can study stability of sheaves on the $z$-plane. 
The stable objects reproduce the dyonic spectrum of the \sw theory
\cite{sw,bilal} both
in $\goth{S}$ and $\goth{W}$. This is established by considering 
 four different cases \cite{ak}. 
\subsection{Ideal sheaf \& Structure sheaf}
\label{sec:ideal}
The two type-IIB 7-branes, 
with charges $(1,-1)$ and $(1,0)$, 
are represented by 
points on the $z$-plane and 
are situated at $z=\pm 1$ 
respectively \cite{asen2}. 
The dyons that become massless on the MS-line 
are wont to stand on them provided they possess matching charges. Indeed, 
restricting our discussions to the upper half plane of \figref{msl}
let us consider the pull-back to ${\mathcal{M}}$ 
of the structure sheaf $\sheaf{O}_{\c}$ and
the ideal sheaf $\sheaf{I}_p \cong \sheaf{O}_{\c}(-1)$, $p$ being the class of a point in
$\cal C$. From the rank and degree of $\sheaf{I}_p$ and $\sheaf{O}_{\cal C}$
we remark that they are the dyonic states with charges
$(1,-1)$ and $(1,0)$, respectively. Their central
charges are 
\begin{equation}
Z(\ipull\sheaf{I}_p) = \vp_D -\vp,
\qquad 
Z(\sheaf{O}_{\mathcal{M}}) = \vp_D,
\end{equation} 
vanishing at $z=-1$ and $z=1$ respectively, which can be checked by explicit
computation of the periods in the lower-half of the 
$z$-plane, $\im z < 0$. In the upper half, on the other hand, the state corresponding
to $\sheaf{O}_{\c}(1)$, with central charge
\begin{equation} 
Z(\ipull\sheaf{O}_{\c}(1)) = \vp_D +\vp
\end{equation} 
becomes massless at $z=-1$. Thus, depending on whether the string approaches 
$z=-1$ from above or below the real axis, it is to be interpreted as
$\sheaf{O}_{\c}(1)$ or $\sheaf{I}_p$, respectively. 
For the former, we shall mark the $z=-1$ point with $\sheaf{O}_{\c}(1)$.
Recalling that the magnitude of the central charge 
gives the mass of the state corresponding to a sheaf, these are the two states
that become massless on the MS-line. Moreover, the charges match with the 
charges of the 7-branes. The sheaves are marked in \figref{msl}.

An $(m,n)$ string corresponding to a dyon may emanate from either of these and end on
the D3-brane probe with any permitted charge, situated at an arbitrary point 
in the $z$-plane, after performing sufficient charge-conserving gymnastics as we
shall discuss in the next section. 
\begin{figure}[h]
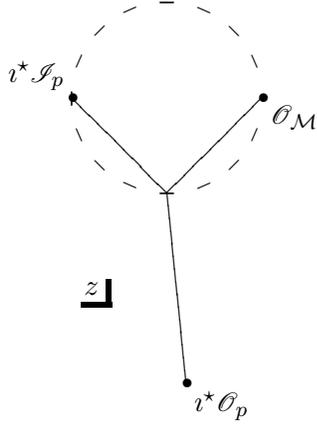

$$\xy
0,/r3pc/:
*\dir{*}="a",*+!DR{\ipull\sheaf{I}_p};
,p+(.2,-2)*+{z}*\frm<2pt>{,},
,p+(2,0)*\dir{*}="b"*+!UL{  \sheaf{O}_{\mathcal{M}}};
,p+(-.8,-3)*\dir{*}="d"*+!UL{ \ipull \sheaf{O}_p},
,p+(-1,-1)*\dir{}="c",
"a";"b",{\ellipse_{--}},
"a";"b",{\ellipse^{--}},
"a";"c", **\crvs{-},
"b";"c", **\crvs{-},
"c";"d", **\crvs{-},
\endxy
$$\caption{The W-boson}
\label{3prong}
\end{figure}
\subsection{Torsion sheaf}
\label{sec:torsion}
The pull-back of the torsion sheaf $\sheaf{O}_p$ is identified with the 
W-boson of the \sw theory corresponding to the dyonic state $(0,1)$.
In the type-IIB probe picture the charge matches with that of the 
D3-brane probe. Its central charge is 
\begin{equation} 
Z(\ipull\sheaf{O}_p) = \vp.
\end{equation} 
It is $\Pi$-stable in the weak coupling region $\goth{W}$
but in the strong coupling region $\goth{S}$ it decays into components according the
the triangle
\begin{equation}
\label{junction:op}
\xymatrix{
\sheaf{I}_p \ar[r] & \sheaf{O}_{\c} \ar[r] & \sheaf{O}_p \ar[r] &
\sheaf{I}_p[1]
}.
\end{equation} 
Thus, a string emanating from the D3-brane 
in $\goth{W}$ may bifurcate on the MS-line into $\sheaf{I}_p$ and 
$\sheaf{O}_{\c}$, which then end on the two 7-branes as discussed above in
\S\S\ref{sec:ideal}.
as indicated in \figref{3prong}.
This provides our first example of a three-string junction \cite{bf1}.
\subsection{Invertible sheaves on $\c$}
The invertible sheaves on $\cal C$, namely $\sheaf{O}_{\cal C}(n)$, $n>0$, correspond
to the dyonic states $(1,n)$ in the \sw theory.
The central charge of this dyon is 
\begin{equation}
Z\big(\ipull\sheaf{O}_{\c}(n)\big)  = \vp_D+ n\vp.
\end{equation} 
These sheaves are stable in the region $\goth{W}$, while decay in the region
$\goth{S}$ into components if $n>1$ \cite{ak}.
Thus these states can be interpreted as string junctions similar to the W-boson
discussed above, but with more than one string in some of the prongs in general. 
However, depending on the position of the D3-brane in the $z$-plane
an $(1,n)$ dyon can also be interpreted, using a different projective resolution, as 
a string spiraling the MS-line in the weak coupling region
$\goth{W}$ before ending on one of the 7-branes at $z=\pm 1$.
We shall come back to these states in more detail in \S\ref{sec:spiral}.
\subsection{Higher rank sheaves}
\label{sec:hi}
\begin{figure}[h]
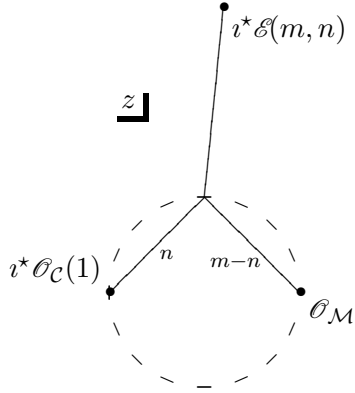

\begin{center}
$$\xy
0,/r3pc/:
*\dir{*}="a",*+!DR{\ipull\sheaf{O}_{\cal C}(1)};
,p+(.2,2)*+{z}*\frm<2pt>{,},
,p+(2,0)*\dir{*}="b"*+!UL{  \sheaf{O}_{\mathcal{M}}};
,p+(-.8,3)*\dir{*}="d"*+!UL{\ipull\sheaf{E}(m,n)},
,p+(-1,1)*\dir{}="c",
"a";"b",{\ellipse_{--}},
"a";"b",{\ellipse^{--}},
"a";"c", **\crvs{-} ? *^!/5pt/{\scriptstyle n},
"b";"c", **\crvs{-} ? *_!/9pt/{\scriptstyle m-n},
"c";"d", **\crvs{-},
\endxy $$
\caption{Higher rank sheaves as junctions}
\label{fig:high}
\end{center}
\end{figure}

A sheaf $\sheaf{E}(m,n)$ with rank $m$ and degree $n$, with $m>1$ 
on $\cp{1}$ splits as 
$\sheaf{E}(m,n)=\sheaf{O}_{\c}(n)\oplus\sheaf{O}_{\c}^{\oplus m-1}$. 
Such states decay on the MS-line according to 
\begin{equation}
\label{high}
\begin{split}
E(m,n) &\rt \sheaf{O}_{\c}(1) + (n-1)\sheaf{O}_p + (m-1)\sheaf{O}_{\c} \\
&\rt \sheaf{O}_{\c}(1) + (n-1) \sheaf{O}_{\c}(1) - (n-1) \sheaf{O}_{\c} + (m-1)
\sheaf{O}_{\c} \\
&\rt n \sheaf{O}_{\c}(1) + (m-n) \sheaf{O}_{\c},
\end{split} 
\end{equation} 
where we used the notation $ n \sheaf{E} = \sheaf{E}^{\oplus n}$ for a sheaf $\sheaf{E}$.
The decay process can be deduced as follows. On the MS-line $\sheaf{O}_{\c}(n)$
decays to $\sheaf{O}_{\c}(1)$ and $\sheaf{O}_p^{\oplus(n-1)}$ \cite{ak}, since
$\sheaf{O}_{\c}(n)$ with $n>1$ are not stable in $\goth{S}$. Hence the
first line in \eq{high}. But nor is $\sheaf{O}_p$ stable in $\goth{S}$. It decays
into $\sheaf{O}_{\c}(1)$ and $\sheaf{O}_{\c}$. Combining this with the earlier decay
process we deduce the decay \eq{high}. Both the products in the last step are stable
in $\goth{S}$.
These states are thus represented as junctions with more than one strings connecting
the vertex to the $z=\pm 1$ points as in \figref{fig:high}.
To summarise, the stable sheaves in the weak coupling region $\goth{W}$ are
$\sheaf{O}_p$ corresponding to the W-boson with dyon charge $(0,1)$ and the invertible sheaves
$\sheaf{O}_{\cal C}(n)$ with dyon charge $(1,n)$ for  $n\geq 0$. 
In the strong coupling region $\goth{S}$, however, the
stable sheaves are the structure sheaf $\sheaf{O}_{\cal C}$ with dyon charge $(1,0)$ 
and the invertible sheaf $\sheaf{O}_{\cal C}(1)$ with dyon charge $(1,1)$. This
corresponds to the spectra of the \sw theory in these two regions \cite{sw,bilal,ak}. 

Let us close this section with a short note on the large volume monodromy
transformation of the sheaves. The large volume monodromy transformation of an object
corresponds to tensoring the object in $\dercat{\c}$ with $\sheaf{O}_{\c}(1)$, which
can be looked upon as a Fourier-Mukai transform\cite{doug4}.
Since coordinate of the \sw moduli space z is related to the type IIA moduli space through the relation $x_1=\frac{1}{z^2}$ 
\cite{ak}, a rotation around $z=0$ in the $z$-plane would correspond to tensoring by $\sheaf{O}_{\c}(-2)$ in the K\"ahler moduli space of $\cal M$.
\section{Complexes, spirals and junctions}
\label{sec:spiral}
We already encountered an example of a three-pronged junction in
\S\S\ref{sec:torsion},
where the dyon corresponding to the sheaf $\ipull\sheaf{O}_p$ was looked upon 
as a string junction bifurcating into two prongs on the MS-line. In this
section we concern ourselves with the interpretation of the invertible sheaves as
spiral strings and junctions. In many of the instances discussed in this section some
of the prongs of a junction will have more than one strings in it, as encountered in
\S\S\ref{sec:hi}.
Since we study the possible decays of an invertible
sheaf $\sheaf{O}_{\cal C}(n)$, and its interpretation as junctions or spiral strings,
we consider projective resolutions of it, which may replace the sheaf in the derived
category. One such resolution which we shall use repeatedly 
is provided by the short exact sequence
\begin{equation}
\label{hn}
\xymatrix{
0\ar[r] & (n-k-1) \sheaf{O}(k)\ar[r] & (n-k)\sheaf{O}(k+1)\ar[r] & \sheaf{O}(n)\ar[r]
& 0
},
\end{equation} 
for $n> k+1$, $n$ and $k$ being integers.
In the above equation as well as in the rest of this section all sheaves are 
on $\cal C$ with their pull-backs assumed to furnish the branes on the \cy
manifold $\cal M$.
Hence we drop the subscript and $\ipull$ as well. 
Let us present a few examples first, which we shall generalise then.
\subsection{Examples}
\begin{figure}[h]
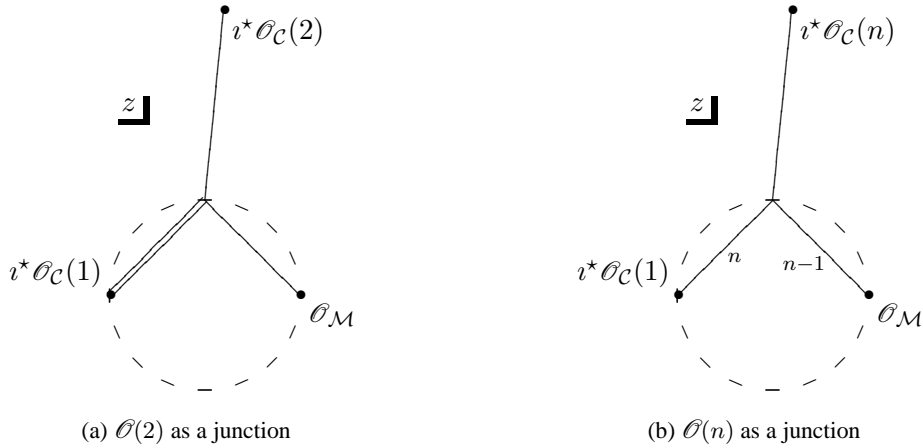

\begin{center}
\subfigure[$\sheaf{O}(2)$ as a junction]{
\xy
0,/r3pc/:
*\dir{*}="a",*+!DR{\ipull\sheaf{O}_{\cal C}(1)};
,p+(.2,2)*+{z}*\frm<2pt>{,},
,p+(2,0)*\dir{*}="b"*+!UL{\sheaf{O}_{\mathcal{M}}};
,p+(-.8,3)*\dir{*}="d"*+!UL{\ipull \sheaf{O}_{\cal C}(2)},
,p+(-1,1)*\dir{}="c",
"a";"b",{\ellipse_{--}},
"a";"b",{\ellipse^{--}},
"a";"c", **\crvs{=},
"b";"c", **\crvs{-},
"c";"d", **\crvs{-},
\endxy
\label{fig:o2}
}
\hskip 1in
\subfigure[$\sheaf{O}(n)$ as a junction]{
\xy
0,/r3pc/:
*\dir{*}="a",*+!DR{\ipull\sheaf{O}_{\cal C}(1)};
,p+(.2,2)*+{z}*\frm<2pt>{,},
,p+(2,0)*\dir{*}="b"*+!UL{  \sheaf{O}_{\mathcal{M}}};
,p+(-.8,3)*\dir{*}="d"*+!UL{\ipull \sheaf{O}_{\cal C}(n)},
,p+(-1,1)*\dir{}="c",
"a";"b",{\ellipse_{--}},
"a";"b",{\ellipse^{--}},
"a";"c", **\crvs{-} ? *^!/5pt/{\scriptstyle n},
"b";"c", **\crvs{-} ? *_!/9pt/{\scriptstyle n-1},
"c";"d", **\crvs{-},
\endxy
\label{fig:on}
}
\caption{Junctions}
\end{center}
\end{figure}

\begin{figure}[h]
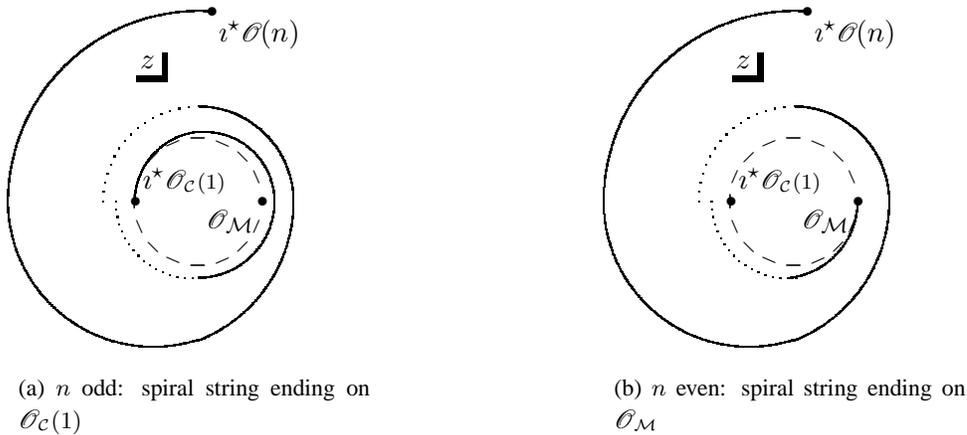

\begin{center}
\subfigure[$n$ odd: spiral string ending on $\sheaf{O}_{\cal C}(1)$]{%
$$\xy
0,/r2pc/:
*\dir{*}="a",*+!DL{\ipull\scriptstyle\sheaf{O}_{\c}(1)};
,p+(.2,2.2)*+{z}*\frm<2pt>{,},
,p+(2,0)*\dir{*}="b"*+!UR{\sheaf{O}_{\mathcal{M}}};
,p+(-.8,3)*\dir{*}="d"*+!UL{\ipull\sheaf{O}(n)}
,p+(-1,1)*\dir{}="c",
,"a";"b",{\ellipse^{--}},
,"a";"b",{\ellipse_{--}},
,p+(-2,0)*\dir{}="m",
,p+(1,-2.2)*\dir{}="n",
,p+(2.5,0)*\dir{}="o",
,p+(1,1.5)*\dir{}="p",
,p+(-.5,0)*\dir{}="q",
,p+(-.3,0)*\dir{}="w",
,p+(1,-1.2)*\dir{}="r",
,p+(2.2,0)*\dir{}="s",
,p+(1,1.1)*\dir{}="t",
,"m";"d",{\ellipse_{-}},
,"m";"n",{\ellipse^{-}},
,"o";"p",{\ellipse^{-}},
,"o";"n",{\ellipse_{-}},
,"q";"p",{\ellipse_{.}},
,"w";"r",{\ellipse^{.}},
,"s";"r",{\ellipse_{-}},
,"s";"t",{\ellipse^{-}},
,"a";"t",{\ellipse_{-}},
\endxy
$$
\label{spiral:o1}
}%
\hskip 1in
\subfigure[$n$ even: spiral string ending on $\sheaf{O}_{\mathcal{M}}$]{%
$$\xy
0,/r2pc/:
*\dir{*}="a",*+!DL{\ipull\scriptstyle\sheaf{O}_{\c}(1)};
,p+(.2,2.2)*+{z}*\frm<2pt>{,},
,p+(2,0)*\dir{*}="b"*+!UR{\sheaf{O}_{\mathcal{M}}};
,p+(-.8,3)*\dir{*}="d"*+!UL{\ipull\sheaf{O}(n)}
,p+(-1,1)*\dir{}="c",
,"a";"b",{\ellipse^{--}},
,"a";"b",{\ellipse_{--}},
,p+(-2,0)*\dir{}="m",
,p+(1,-2.2)*\dir{}="n",
,p+(2.5,0)*\dir{}="o",
,p+(1,1.5)*\dir{}="p",
,p+(-.5,0)*\dir{}="q",
,p+(-.3,0)*\dir{}="w",
,p+(1,-1.2)*\dir{}="r",
,"m";"d",{\ellipse_{-}},
,"m";"n",{\ellipse^{-}},
,"o";"p",{\ellipse^{-}},
,"o";"n",{\ellipse_{-}},
,"q";"p",{\ellipse_{.}},
,"w";"r",{\ellipse^{.}},
,"b";"r",{\ellipse_{-}},
\endxy
$$
\label{spiral:o}
}
\end{center}
\caption{Spiral strings ending on different branes}
\label{fig:spiral}
\end{figure}

\begin{example}
First, let us consider the invertible sheaf $\sheaf{O}(2)$ of degree $2$.
Putting $k=0$ and $n=2$ in \eq{hn} we get
\begin{equation}
\label{seso2}
\xymatrix{
0\ar[r] & \sheaf{O}\ar[r] & 2\sheaf{O}(1)\ar[r] & \sheaf{O}(2)\ar[r]
& 0
}.
\end{equation} 
In order to study the stability of $\sheaf{O}(2)$ against decaying into the
components we compute the difference of their grades, 
\begin{equation}
\label{eq:grades}
\varphi\big(2\sheaf{O}(1)\big) - \varphi (\sheaf{O}) -1 
= -1 -\frac{1}{\pi}\arg\left(1+\vp/\vp_D\right).
\end{equation} 
As we cross the MS-line the difference increases through zero. That is,
$\sheaf{O}(2)$ decays into $\sheaf{O}$ and two copies of $\sheaf{O}(1)$ on the 
MS-line (We are not explicitly differentiating between brane and antibrane as 
that is obvious from the grading). The above sequence
is represented as a junction with three prongs. 
However, now there are two strings that can end on the 7-brane at $z=-1$ 
corresponding to the two copies of $\sheaf{O}(1)$ in \eq{seso2} so that the 
corresponding prong has two strings as shown in 
\figref{fig:o2}.  
\end{example}
\begin{example}
Let us now consider the degree three invertible sheaf $\sheaf{O}(3)$. 
Putting $k=0$ and $n=3$ in \eq{hn} we
derive 
\begin{equation} 
\label{o3prong}
\xymatrix{
0\ar[r] & 2\sheaf{O}\ar[r] & 3\sheaf{O}(1)\ar[r] & \sheaf{O}(3)\ar[r]
& 0
}.
\end{equation} 
On the MS-line $\sheaf{O}(3)$ decays into the components, three copies of
$\sheaf{O}(3)$ and two copies of $\sheaf{O}$, governed again by 
\eq{eq:grades}. Thus this corresponds again
to a three-pronged junction. Three strings join the junction to the $z=-1$
point on the MS-line and two join the vertex to $z=1$, as in \figref{fig:on} for
$n=3$. 

This case presents another possibility, however.
Putting $n=3$ and $k=1$ in \eq{hn} we get another projective resolution of
$\sheaf{O}(3)$ given by the short exact sequence 
\begin{equation} 
\label{o3spiral}
\xymatrix{
0\ar[r] & \sheaf{O}(1)\ar[r] & 2\sheaf{O}(2)\ar[r] & \sheaf{O}(3)\ar[r]
& 0
}.
\end{equation} 
Now, in the derived category, an object may be replaced by its 
projective resolution, without altering its cohomology. Furthermore,
the sheaf $\sheaf{O}(3)$ is $\Pi$-stable with respect to decaying
into the other two sheaves appearing in the resolution \eq{o3spiral}. 
Hence, we can trade the resolution 
$\xymatrix{0\ar[r]&\sheaf{O}(1)\ar[r]&2\sheaf{O}(2)}$ 
for the dyon $\sheaf{O}(3)$.
The string emanating from the D3-brane at an
arbitrary $z$, corresponding to $\sheaf{O}(3)$ ends as 
$\sheaf{O}(1)$ on the 7-brane at $z=-1$. It goes around the MS-line 
once before doing so, thereby
reducing its charge by undergoing a large volume monodromy 
transformation, corresponding to being tensored by $\sheaf{O}(-2)$, from $(1,3)$ of 
$\sheaf{O}(3)$ to $(1,1)$ of $\sheaf{O}(1)$. Hence,
this corresponds to a spiral, as shown in \figref{spiral:o1}.
Let us point out that the sheaves in the resolution are stable only in the weak
coupling region $\goth{W}$. Hence the spiral never crosses the MS-line.
\end{example}
\begin{example}
We consider the sheaf $\sheaf{O}(4)$ with degree four next. 
It appears in the short exact sequence 
\begin{equation}
\xymatrix{
0\ar[r]& 3\sheaf{O}\ar[r]& 4\sheaf{O}(1)\ar[r]&\sheaf{O}(4)\ar[r]&0
},
\end{equation} 
obtained from \eq{hn} by putting $n=4$ and $k=0$.
It decays into the components on the MS-line, governed once
again by \eq{eq:grades}.
As before, then, this corresponds to a junction as in 
\figref{fig:on} with four strings
connecting the vertex to the 7-brane at 
$z=-1$ and three connecting it to the 7-brane
at $z=1$.

In order to see the spiral string corresponding to this dyon we 
consider another projective resolution of $\sheaf{O}(4)$. 
Putting $n=4$ and $k=n-2$ in \eq{hn} we obtain the short exact sequence
\begin{equation}
\label{o4spiral}
\xymatrix{
0\ar[r] & \sheaf{O}(2)\ar[r] & 2\sheaf{O}(3) \ar[r] & \sheaf{O}(4)\ar[r] & 0
}. 
\end{equation} 
This, as it is can not be interpreted as a spiral, nor as a junction, since
$\sheaf{O}(2)$ is not stable inside the MS-line. The resolution corresponding to a
spiral must end on the left with either $\sheaf{O}$ or $\sheaf{O}(1)$, in order to be
permitted to end on either  7-brane. In order to construct such a resolution 
we splice the sequence \eq{o4spiral} with \eq{seso2} to form the Yoneda composite 
\begin{equation} 
\label{o4y}
\xymatrix{
0\ar[r]& \sheaf{O}\ar[r]& 2\sheaf{O}(1)\ar[r]&
2\sheaf{O}(3)\ar[r]&\sheaf{O}(4)\ar[r]&0
}.
\end{equation} 
The sheaves participating in this exact sequence are stable in $\goth{W}$. 
Hence, as before, we can trade in the projective resolution 
$\xymatrix{0\ar[r]& \sheaf{O}\ar[r]& 2\sheaf{O}(1)\ar[r]&
2\sheaf{O}(3)}$
for the dyon $\sheaf{O}(4)$ in $\dercat{\c}$. Thus the
string corresponding to the dyon $\sheaf{O}(4)$ 
emanating from the D3-brane in $\goth{W}$ ends on the 7-brane at $z=1$ as
$\sheaf{O}$.
The string spirals the MS-line twice in $\goth{W}$ and never going beyond this region, 
thus reducing charges through large
volume monodromy from $(1,4)$ to $(1,0)$, 
before ending on $\sheaf{O}$ \cite{bf2}, as in \figref{spiral:o}. 
\end{example}
\begin{example}
Let us now consider the degree five sheaf
$\sheaf{O}(5)$ as our final example before we generalise these 
considerations. 
Putting $n=5$ and $k=0$ in \eq{hn} we derive the short exact sequence
\begin{equation}
\label{o5junction} 
\xymatrix{
0\ar[r] & 4\sheaf{O}\ar[r]&5\sheaf{O}(1)\ar[r]&\sheaf{O}(5)\ar[r]&0
}.
\end{equation} 
The dyon $\sheaf{O}(5)$ decays, one more time, 
to the components on the MS-line governed
by \eq{eq:grades} and hence interpreted, as before, as a three-pronged junction 
as in  \figref{fig:on} with $n=5$ with five strings between the vertex and the
7-brane at $z=-1$ and four between the vertex and the 7-brane at $z=1$.

The spiral string, once again, is obtained by considering a different projective
resolution of $\sheaf{O}(5)$ in terms of sheaves stable in $\goth{W}$.  
Putting $n=5$ and $k=n-2=3$ in
\eq{hn} we obtain
\begin{equation}
\label{eq:o5}
\xymatrix{
0\ar[r] & \sheaf{O}(3)\ar[r]& 2\sheaf{O}(4)\ar[r]&\sheaf{O}(5)\ar[r]&0
} .
\end{equation} 
Since the sheaves $\sheaf{O}(4)$ and  $\sheaf{O}(3)$ are
not stable in the strong coupling region, this sequence can not be interpreted as a
junction. We need to find resolutions ending with either $\sheaf{O}$ or
$\sheaf{O}(1)$ in order to obtain a spiral string.
Splicing the short exact sequence \eq{eq:o5}
with \eq{o3spiral} we derive the exact sequence
\begin{equation} 
\label{o5spiral}
\xymatrix{
0\ar[r] & \sheaf{O}(1)\ar[r]& 2\sheaf{O}(2)\ar[r]&
2\sheaf{O}(4)\ar[r]&\sheaf{O}(5)\ar[r]&0
}.
\end{equation} 
The sheaves in this exact sequence are stable in $\goth{W}$ and hence
$\sheaf{O}(5)$ is replaced with its resolution in $\dercat{\c}$. This time, however,
the string ends as $\sheaf{O}(1)$ on the 7-brane at $z=-1$, as in the case of
$\sheaf{O}(3)$. The right amount of reduction of charges  
by large volume monodromy requires, then, 
that the string spirals around the MS-line four times before ending on the 7-brane.
\end{example}
\subsection{General formulas \& transitions}
These considerations exemplify a general situation to which we now turn.
The strategy is exactly the same as we have discussed in the context of the above
examples. Namely, to obtain a three-pronged junction we need a short exact sequence
which provides a resolution of a degree $n$ invertible sheaf $\sheaf{O}(n)$ in terms
of the sheaves $\sheaf{O}$ and $\sheaf{O}(1)$. Then, a spiral string is obtained as
another resolution ending on the left with either $\sheaf{O}$ or $\sheaf{O}(1)$, with
a sequence of sheaves in-between, all of which are stable in the weak coupling region
$\goth{W}$. The number of windings of the spiral around the MS-line is obtained by
requiring charge conservation and remarking that the degree of a sheaf reduces by $2$
under the large volume monodromy transformation.
Finally, we study the transitions between the two types of resolutions.

In generalising the examples along the lines described 
above let us first note the projective resolution of 
an invertible sheaf $\sheaf{O}(n)$ of degree $n$ given by the short
exact sequence
\begin{equation}
\label{ses:gen}
\xymatrix{
0\ar[r] & (n-1)\sheaf{O}\ar[r] & n\sheaf{O}(1)\ar[r]&\sheaf{O}(n)\ar[r]&0
},
\end{equation} 
which is obtained by putting $k=0$ in \eq{hn}. The sheaf $\sheaf{O}(n)$ breaks
into the components on the MS-line, the decay being governed by \eq{eq:grades}. 
Hence, this is interpreted as a junction as in \figref{fig:on} with $n$ strings
connecting the $z=-1$ point and $n-1$ strings connecting the $z=1$ point to the
vertex of the junction.

The resolutions corresponding to the spiral strings are obtained by 
forming Yoneda composites of complexes iteratively, as in the examples above.
By putting $k=n-2$ in \eq{hn} we derive 
\begin{equation}
\label{ses:gen2}
\xymatrix{
0\ar[r]&\sheaf{O}(n-2)\ar[r]&2\sheaf{O}(n-1)\ar[r]& \sheaf{O}(n)\ar[r]&0
}.
\end{equation} 
For $n>2$ the sheaves on the left are not stable in the strong coupling region
$\goth{S}$ rendering the sequence inappropriate for being interpreted as a junction.
In order to obtain the projective resolutions suitable for our purpose
we proceed iteratively, by finding a resolution of the leftmost sheaf 
in the above sequence \eq{ses:gen2}, namely, $\sheaf{O}(n-2)$
by replacing $n$ by $n-2$ in \eq{hn} and putting $k=(n-2)-2$. 
This yields 
\begin{equation}
\label{ses:int}
\xymatrix{
0\ar[r]&\sheaf{O}(n-4)\ar[r]&2\sheaf{O}(n-3)\ar[r]& \sheaf{O}(n-2)\ar[r]&0
}.
\end{equation} 
If $n\geq 4$, then this sequence again fails to correspond to a junction. 
We splice it with the sequence \eq{ses:gen2},
deriving another projective resolution of $\sheaf{O}(n)$ as
\begin{equation} 
\xymatrix{
0\ar[r]&\sheaf{O}(n-4)\ar[r]&2\sheaf{O}(n-3)\ar[r]&2\sheaf{O}(n-1)\ar[r]&
\sheaf{O}(n)\ar[r]&0.
}
\end{equation}   
Continuing this and splicing the resulting resolutions to form
Yoneda composites iteratively until we have a resolution ending of the left 
with either a single $\sheaf{O}$ or $\sheaf{O}(1)$, 
we arrive at a projective resolution of
$\sheaf{O}(n)$ as
\begin{equation}
\label{res:even}
\xymatrix{
0\ar[r]&\sheaf{O}\ar[r]&2\sheaf{O}(1)\ar[r]&
\ar[r]\cdots\ar[r]&2\sheaf{O}(n-3)\ar[r]&2\sheaf{O}(n-1)\ar[r]&\sheaf{O}(n)\ar[r]&0
}
\end{equation} 
if $n$ is an even integer, and
\begin{equation} 
\label{res:odd}
\xymatrix{
0\ar[r]&\sheaf{O}(1)\ar[r]&2\sheaf{O}(2)\ar[r]&
\ar[r]\cdots\ar[r]&2\sheaf{O}(n-3)\ar[r]&2\sheaf{O}(n-1)\ar[r]&\sheaf{O}(n)\ar[r]&0
}
\end{equation} 
if $n$ is odd. 
Since all the sheaves participating in the resolution of 
$\sheaf{O}(n)$ are  stable $\goth{W}$, we can trade in the 
projective resolution for $\sheaf{O}(n)$ in $\dercat{\c}$. As in the
examples above, the corresponding spiral string ends at $z=1$ as $\sheaf{O}$ if $n$
is even and at $z=-1$ as $\sheaf{O}(1)$ if $n$ is odd. Since the large volume
monodromy acts on a sheaf by tensoring with $\sheaf{O}(-2)$, the string
spirals around $n/2$ times if $n$ is even and $(n-1)/2$ times if $n$ is odd, before
ending up on the MS-line, in keeping with earlier findings \cite{bf2}. 

Let us now discuss the transitions between the spiral strings and the junctions. 
So far we discussed the interpretation of the same dyon state $(1,n)$
with $n>1$ as either a three-pronged junction or a spiral string ending at $z=\pm 1$.
But at a certain point of the moduli space we expect a unique
description of the dyons \cite{hauer}. 
This means, in the moduli space the dyon is described by the junction in a certain
region and by a spiral string in some other. On the boundary of these two regions
there must be transitions between them \cite{bf2}.

Now, the replacement of an object $C$ in $\dercat{\cal M}$ by its projective
resolution, as given by \eq{abc} while always possible mathematically,  is but 
physically meaningful when the charges of the branes 
are properly aligned to preserve some SUSY \cite{doug1}, as is encoded in the
condition for $\Pi$-stability \eq{piphi}. This requires the grades of $B$ and $C$ of
\eq{abc} to be equal. 
Hence, in order to determine which of the resolutions is to be chosen to replace
$\sheaf{O}(n)$, among the two possibilities discussed above, we need to compare the
grades of the objects in the resolutions. The appropriate resolution 
at a certain point in the $z$-plane is chosen depending on which of the resolutions
satisfy \eq{piphi}. In order to see this explicitly, let us first 
rewrite the sequences \eq{res:even} and
\eq{res:odd} as 
\begin{gather}
\label{beven}
\xymatrix{
0\ar[r] & \sheaf{O}\ar[r]&\sheaf{B}\ar[r]&\sheaf{O}(n)\ar[r]&0,
}
\intertext{and}
\label{bodd}
\xymatrix{
0\ar[r] & \sheaf{O}(1)\ar[r]&\sheaf{B}\ar[r]&\sheaf{O}(n)\ar[r]&0,
}
\end{gather} 
respectively by denoting the intermediate complex by $\sheaf{B}$. 
These correspond to the spirals as discussed before. For all of 
these
resolutions the grade of the complex $\sheaf{B}$ is  
\begin{equation} 
\varphi (\sheaf{B}) = -1, 
\end{equation} 
calculated from its Chern character $\ch{\sheaf{B}} = (0,n)$, $n>1$, using the large volume
expression \eq{ze:ch}. 
This is to be contrasted with the situation with the short exact sequence
\eq{ses:gen}. The middle term is
$n\sheaf{O}(1)$, with grade 
\begin{equation} 
\varphi\big(n\sheaf{O}(1)\big)=-3/2
\end{equation} 
to the leading order in $\goth{W}$. 
Now, the replacement of the sheaf $\sheaf{O}(n)$ in the derived category
$\dercat{\c}$ by a projective resolution as in \eq{abc} with $C=\sheaf{O}(n)$
is physically proper if its grade equals that of $B$, as given by \eq{piphi}.
For the spirals \eq{beven} and \eq{bodd}, $\phi (B)=-1$, while for the three-pronged
junction \eq{ses:gen} we have $\phi (B)=-3/2$. 
Thus, depending on whether the grade of $\sheaf{O}(n)$ is $-3/2$ or $-1$, the
resolution appropriate for it to be replaced with is \eq{ses:gen} or either of
\eq{res:even} or \eq{res:odd}, if $n$ is even or odd, respectively. 
This change of grade occurs as the dyon $(1,n)$ crosses the 
line $\arg (z) = \pi/2$ in the moduli space. Hence, as the dyon crosses this line,
a spiral string
corresponding to $\sheaf{O}(n)$, given by either of \eq{res:even} and 
\eq{res:odd} makes transition into the junction given by \eq{ses:gen}, 
which agrees with earlier results \cite{bf2}. These cases exhaust the possibilities 
that come from \eq{hn}. Other values of $k$ do not lead to qualitatively new resolution.
\section{Conclusion}
To conclude, in this article we have considered the dyons in the \sw theory from 
the point of view of B-branes wrapping holomorphic cycles in a \cy manifold $\cal M$
in the 
geometrical engineering limit. Such B-branes correspond to the derived category of
coherent sheaves on $\cal M$. In the geometric engineering limit this category is
assumed to be generated by the coherent sheaves on the projective line, $\cp{1}$,
the sole compact survivor of the limit \cite{ak}. The $\Pi$-stable sheaves in this
derived category reproduce the dyons in the \sw theory \cite{ak}. The notion of the
derived category stems from the need to identify an object of an abelian category
with its resolutions. By introducing a grade in the derived category, the criterion of
$\Pi$-stability provides a physical meaning to such replacements. The replacement is
physically meaningful provided the grades of the objects participating in the
resolution obey certain conditions \eq{piphi}. A dyon of the \sw theory decays on a
certain marginal stability locus, the MS-line, according to this condition. The
stable dyons on the two sides of the line are different. An estimation of the stable
dyons on the two sides involve distinguished triangles in the triangulated category
underlying $\dercat{\cp{1}}$, which, in turn, correspond to short exact sequences in
the abelian category. These triangles are interpreted as three-pronged junctions,
involving prongs corresponding to strings on both sides of the MS-line
resting on the stable dyons on the line itself. In general the prongs of a junction
are found to have more than one strings in it, consistent with earlier findings
\cite{bf1}. Considering certain other projective
resolutions of the invertible sheaves on $\cp{1}$, involving sheaves which are stable
on only one side of the MS-line, the weak coupling region of the \sw theory, we
show that such resolutions do not correspond to pronged junctions, as they are not
allowed to penetrate the strong-coupling region beyond the MS-line. These resolutions,
obtained by forming Yoneda composites of short exact sequences of sheaves on $\cp{1}$
iteratively, are interpreted as strings spiraling the MS-line in the weak coupling
region, ending finally on  7-branes situated on the MS-line at $z=\pm 1$. Using
the condition \eq{piphi} we then compare the grades of the two resolutions and find
that they may replace a dyon corresponding to an invertible sheaf
$\sheaf{O}_{\cp{1}}(n)$, with $n>1$ in but different regions of the moduli space
spanned by $z$. As the dyon crosses the line $\arg (z) = \pi/2$ in the $z$-plane, a
three-pronged junction takes over a spiral string or vice versa, thus corroborating
earlier results \cite{bf2}. 

These considerations, apart from reproducing the junctions and spirals of the \sw
theory and their transitions, clearly brings out the role of the grade of
$\Pi$-stability in deciding the replacement for an object in the derived category in a
physically meaningful manner. It will be interesting to see if similar considerations
predict such transitions in more complicated situations, for example, for string
junctions corresponding to \sw theories arising from D-branes on del Pezzo surfaces.
Investigation along this line is currently underway. 
\section*{Acknowledgment}
SM and KR would like to thank Ansar Fayyazuddin  and Alok Kumar 
for useful discourses at different
stages of this work. The work of SM is supported by NSF grant PHY--0244801.

\end{document}